# Beyond 100 nm Resolution in 3D Laser Lithography – Post Processing Solutions

G. Seniutinas *[a], A. Weber [a, b], C. Padeste [a], I. Sakellari [c], M. Farsari [c], and C. David [a]

*(a) Paul Scherrer Institut, Villigen PSI, 5232, Switzerland*
*(b) Laboratory for Mesoscopic Systems, Department of Materials, ETH Zurich, 8093 Zurich, Switzerland*
*(c) FORTH, Institute of Electronic Structure and Laser, Heraklion, 71110, Greece*

*\*Corresponding author*
e-mail address: gediminas.seniutinas@psi.ch

**Abstract**

Laser polymerization has emerged as a direct writing technique allowing the fabrication of complex 3D structures with microscale resolution. The technique provides rapid prototyping capabilities for a broad range of applications, but to meet the growing interest in 3D nanoscale structures the resolution limits need to be pushed beyond the 100 nm benchmark, which is challenging in practical implementations. As a possible path towards this goal, a post processing of laser polymerized structures is presented. Precise control of the cross-sectional dimensions of structural elements as well as tuning of an overall size of the entire 3D structure was achieved by combining isotropic plasma etching and pyrolysis. The smallest obtainable feature sizes are mostly limited by the mechanical properties of the polymerized resist and the geometry of 3D structure. Thus the demonstrated post processing steps open new avenues to explore free form 3D structures at the nanoscale.

1. Introduction

Three-dimensional (3D) laser lithography has become an established technique that allows direct writing of arbitrary 3D structures at the microscale. The achievable resolution and versatility make this fabrication approach highly suitable for numerous applications ranging from micro-optical elements [1-4] and 3D templates [5] to micro-scaffolds for cell studies [6-9]. However, the growing interest in nanoscale 3D objects requires sub-100 nm resolution, which still remains demanding in practical applications.

Structuring by 3D laser lithography is based on local curing of a resist with a tightly focused laser beam. The obtainable resolution mainly depends on light-matter interaction and localization of photons, thus these two aspects have been tackled to push the resolution limits. Solutions from the optical microscopy were successfully applied to deliver photons of a wavelength λ into a diffraction-limited spot size of ~λ/2, reaching fundamental limitations of light confinement in the far field. Hence, focal spot sizes of a few hundreds of nanometers can be achieved for lasers operating at visible to near infrared wavelengths. Interestingly, the smallest feature size of the exposed structure may still be smaller than the diffraction limited spot size due to nonlinear light-matter interaction effects and by taking advantage of selected cross-linking thresholds. To tailor the light-matter interaction and efficiently exploit nonlinear effects novel resist materials were developed [10-12]. Their application in multiphoton polymerization [13-15] as well as stimulated emission depletion techniques [16, 17] resulted in achievable resolutions well below 100 nm. However, complicated writing setups and stability issues currently hinder practical applications of free form 3D structuring beyond 100 nm feature sizes and commercially available systems mostly offer writing resolution only in the sub-micrometer range.

An alternative approach to produce well defined nanoscale features is the pyrolysis of laser polymerized structures. A few studies reported that the size of 3D objects can be efficiently reduced by thermal decomposition of a resist in vacuum or inert gas atmosphere [18, 19]. The decomposition results in mass loss of up to 80% while turning the resin into a glassy carbon [19, 20, 21]. This change is accompanied by shrinkage of the structure and reduction of its dimensions. Pyrolysis allows scaling of the entire structure; however, it does not leave much freedom to tune width/thickness of its building blocks.

To add nanoscale precision and flexibility to post processing approaches, we studied the combination of isotropic plasma etching and pyrolysis for fabrication of 3D structures with sub-100 nm feature sizes. Plasma etching provides accurate control of cross-sectional dimensions of structural elements while an overall size of the entire 3D structure is down scaled via pyrolysis.

2. Materials and Fabrication

Two types of resists were used to polymerize test structures. The first one was an organic negative tone photopolymer known under the brand name IP-Dip and containing 60–80% of 2-(hydroxymethyl)-2-[[(1-oxoallyl)oxy]methyl]-1,3-propanediyl diacrylate (also known as Pentaerythritol triacrylate). The resist was developed by Nanoscribe GmbH and serves as an immersion and photosensitive material at the same time. The material (25 g bottle, batch: 1-600-0055) was obtained from Nanoscribe GmbH and used according to the provided standard operation procedures for 3D structuring using a commercially available laser lithography system Photonics Professional GT (Nanoscribe GmbH).

The second type of resist was an organic–inorganic zirconium–silicon hybrid composite (SZ2080 [10]) doped with a quencher 2-(dimethylamino) ethyl methacrylate (DMAEMA) [22, 23]. The resist is composed of an inorganic network and an organic one. The inorganic network consists of Si-O-Si, Si-O-Zr and Zr-O-Zr bonds and is formed during the synthesis of the material prior to photo-polymerization. The organic network is the one formed during photo-polymerization by polymerizing the pendant methacrylate moieties.

The photoresist was produced using the sol-gel method [24]. It consists of methacryloxy-propyl trimethoxysilane (MAPTMS), zirconium propoxide (ZPO, 70% in propanol), and DMAEMA. MAPTMS and DMAEMA were used as the organic photopolymerizable monomers, while ZPO and the alkoxysilane groups of MAPTMS were used to form the inorganic network. In our case, the organic–inorganic hybrid material was synthesized by mixing the above chemicals in the following molar ratios: MAPTMS:ZPO = 8:2, and (MAPTMS+ZPO):DMAEMA = 9:1. Michler's ketone (4,4-bis(diethylamino) benzophenone) used as a photoinitiator was added at 1% w/w concentration to the final solution of MAPTMS and DMAEMA composite. All chemicals were obtained from Sigma-Aldrich and used without further purification. The samples were prepared by drop casting onto 100 μm thick silanized glass substrates and the resulting films were dried in air for several days before photopolymerization.

A Ti:Sapphire femtosecond laser (Femtolasers Fusion, 800 nm, 75 MHz, <20 fs) was used as the light source for the fabrication of three-dimensional microstructures. The complete experimental setup and procedure has been described elsewhere [25]. The laser beam was focused using a high numerical aperture microscope objective lens 100x, N.A. 1.4, Zeiss, Plan Apochromat. The writing speed employed was 5 μm s$^{-1}$ and the peak intensity in the focal spot was ~0.35 TW/cm$^2$. After the completion of the component building process by Direct Laser Writing, the samples were developed for 20 min in a 50:50 solution of 1-propanol/2-propanol and were further rinsed with 2-propanol.

A buckyball structure shown in Figure 1 was selected as a 3D model for our experiments. The geometry of the structure provides high mechanical stability that allows reaching sub-100 nm widths of the composing bars without a structural collapse. The focal spot of the laser is elongated along the optical axis and the voxel has an aspect ratio of around 1:3. To achieve the highest resolution, single pixel lines were exposed, thus composing bars are much thicker along the focal spot (z axis) than in the xy plane.

A tube furnace was used for pyrolysis of laser polymerized buckyballs. The structures were baked at 690 $^o$C in a tube furnace under a constant nitrogen flow at 100 sccm throughout the chamber. At the beginning, the temperature was ramped up to 250 $^o$C and kept at this point for 60 min allowing the remaining solvents in the structures to evaporate. Then the temperature was raised to 350 $^o$C at a ramp rate of 5 $^o$C/min. At this temperature the organic constituents of the resists start to decompose [3, 26], thus the samples were kept for another 60 min at 350 °C to ensure slow mass loss process and minimized distortions. Finally, the temperature was set to 690 °C keeping the ramp rate at the same value and the structures were kept for 60 min at this temperature to decompose the remaining functional carbon groups. The heating was then switched off and the buckyballs left for a few hours to cool down under the nitrogen flow.

Isotropic plasma etching was carried out in a parallel plate etcher. The etcher uses a radio frequency generator operating at 27 MHz to produce plasma inside the chamber. Biasing is negligible due to the parallel plate configuration, thus structures are being etched in a highly isotropic manner. The etching experiments were done in cycles. First the structures were etched for 1 minute in oxygen plasma and then inspected in a high resolution SEM (Zeiss Supra 55VP). After the SEM imaging, the same structures were etched again for 1 minute and the inspection step repeated. This etch-inspect cycle was repeated until noticeable distortions of the structure occurred.

The experiments were done with sets of 2 substrates each having at least 8 buckyball structures to evaluate experimental repeatability. Fabrication and measurement errors are given together with the measured values.

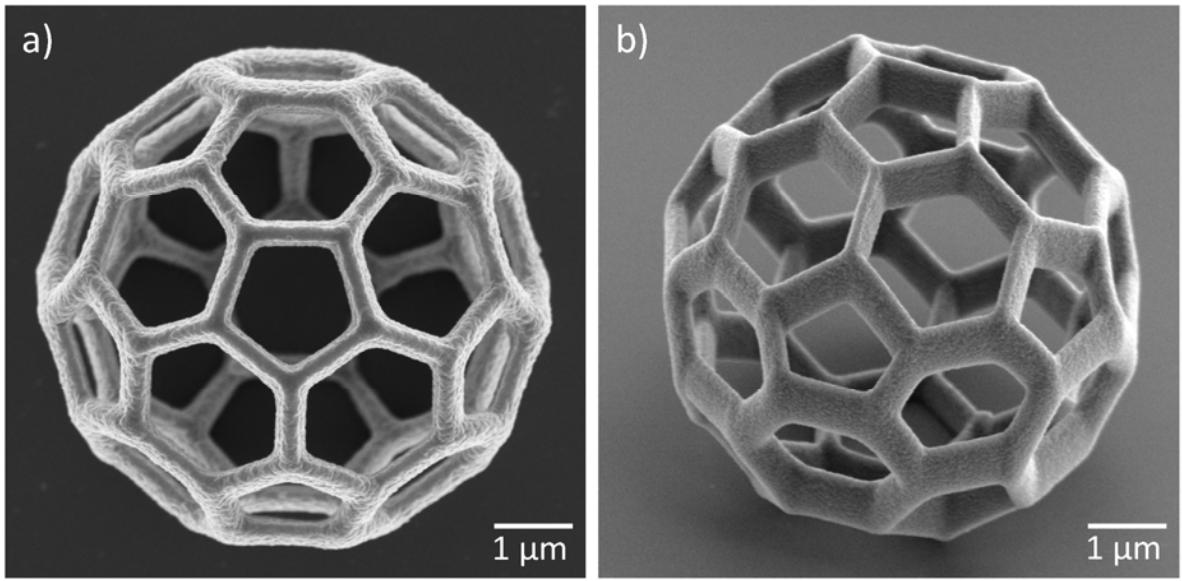

Figure 1: 7 µm diameter laser polymerized buckyball used as a 3D model for experiments. The structure was produced using SZ2080 and imaged before the post processing steps. a) top view; b) 45 degree tilt view showing 1:3 aspect ratio of the composing bars.

# 3. Results and Discussion

## 3.1. *Post Processing: Pyrolysis*

During pyrolysis the resist molecules were decomposed and volatile species were removed from the chamber. Depending on the initial composition of the resist, the remaining elements rearrange to form a glassy carbon [21] (IP-DIP) or zirconium doped silica-carbon mixture [18] (SZ2080). The mass loss and rearrangement of the material cause the structures to shrink. Figure 2a shows an initial $7.0 \pm 0.1$ µm diameter buckyball polymerized using IP-Dip resist. During pyrolysis the dimensions of the resulting structures were reduced to 20-25% of their initial size (Figure 2b). The same pyrolysis procedure was repeated for the buckyballs prepared using SZ2080 photoresist. As the organic content of SZ2080 is much lower, the structures shrunk less, i.e. to 60-70% of their initial size (Figure 3). Presumably, these structures are composed of zirconium doped silica and carbon, but further investigations are needed to determine what part of organic component is remaining after the pyrolysis. Similar shrinkage values for IP-Dip and SZ2080 resists have been reported earlier in [19] and [3] respectively.

As can be noted from Figure 2b, the lower part of the structures which is attached to the substrate shrunk less due to strong adhesion forces. This caused the structures to be distorted and not entirely matching their initial geometry. Such distortions – in a controlled manner – could be used as a lever to produce more complex 3D architectures at the nanoscale. By controlling strong adhesion points and the degree of shrinkage, a distortion pattern can be engineered. One of the examples of high resolution 3D nanostructures made by following this approach is shown in Figure 2c. Here, micro-domes were made by pyrolyzing buckyballs made from IP-Dip resist. It would be extremely challenging to make such structures using just laser polymerization alone due to the required resolution and a highly reflective silicon substrate. Also, during pyrolysis material properties are altered. Initially insulating IP-Dip is converted into a glassy carbon that has high thermal stability and is electrically conductive, thus 3D electrical circuits could be engineered at the nanoscale.

In case deformations are not desired, they can be minimized by including additional structural elements into the main structure [17, 18]. These elements should act as an intermediate link between the primary structure and a substrate reducing a coupling between the two. In our case, the structure was placed on top of a pillar (Figure 4a). This modified design was tested following the same pyrolysis procedures as before and the resulting down scaled structure preserved its initial geometry (Figure 4b).

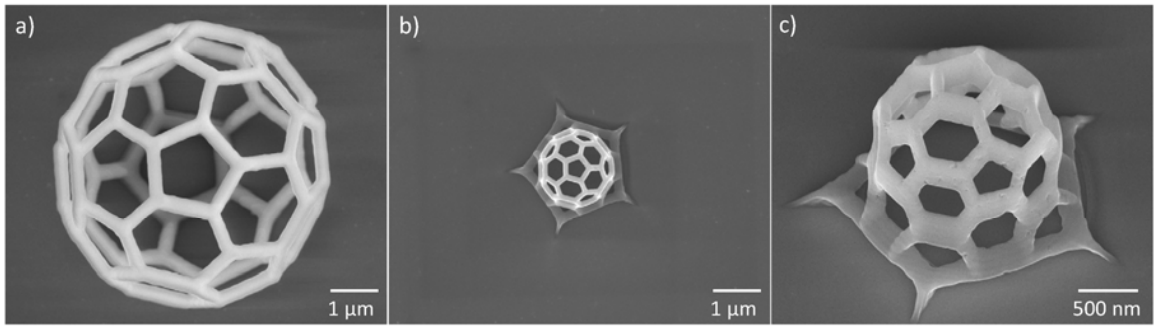

Figure 2: Shrinkage of the polymerized structures due to pyrolysis: a) SEM images of the initial 7 μm diameter buckyball. IP-Dip resist and Nanoscribe system was used to produce the structures on a silicon substrate; b) corresponding structures after the pyrolysis; c) 45 degree tilt view of the pyrolyzed structure. Note: SEM magnification is the same in a) and b).

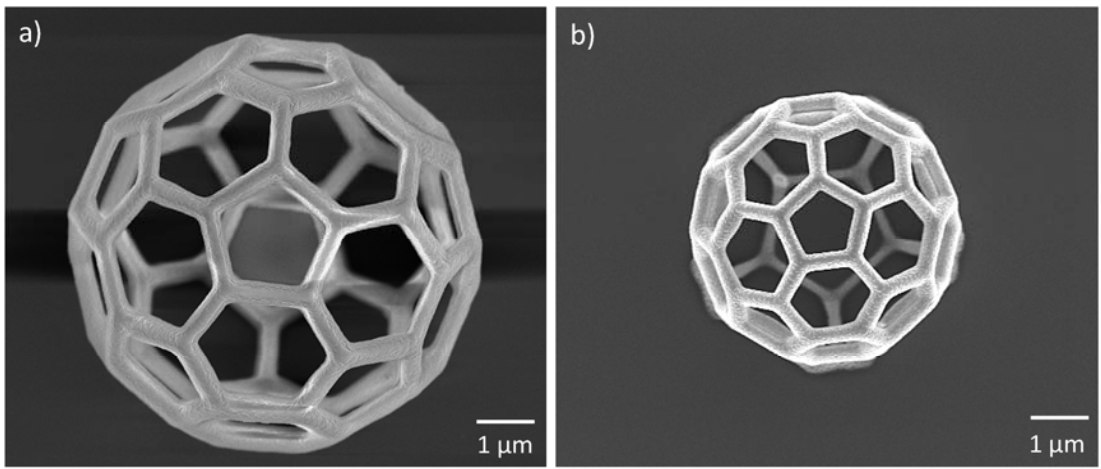

Figure 3: Top view of a buckyball polymerized using SZ2080 photoresist:
a) initial structure; b) after the pyrolysis.

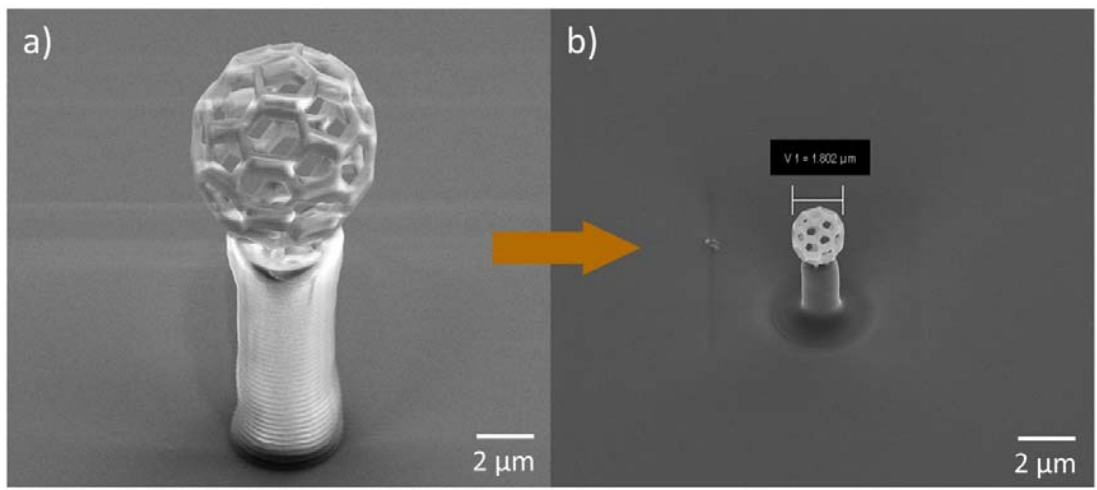

Figure 4: Buckyball on a decoupling pillar: a) initial structure made using IP-Dip; b) structure obtained after the pyrolysis. Note, both images were taken at the same SEM magnification.

## 3.2. Post Processing: Isotropic Plasma Etching

Pyrolysis allows reduction of an overall size of the entire structure. However, it does not provide much freedom of independently changing the lateral dimensions of the composing structural elements. To add a flexibility of tuning the bar width, isotropic plasma etching was investigated as a post-processing step. Both of the investigated resists have some part of an organic component, thus oxygen plasma was used for the etching. Moreover, oxygen plasma also does not etch typical substrate materials, such as silicon or glass, used for 3D laser lithography.

Figures 5a-c show SEM images of the same IP-Dip resist structure before the etching, after 5 and after 7 etching cycles, respectively. The etching rate was measured to be 15 nm/min and it was constant for every etching cycle. The etch rate of exposed SZ2080 resist was highly dependent on the etching cycle, as can be seen from the SEM images in Figures 5d-f. After the 6$^{th}$ cycle, no noticeable etching was observed using oxygen plasma. Presumably, at that point the organic part of the polymer backbone has already been etched away. What remains probably is the inorganic network, which is a glass-like network doped with zirconium. Further thinning of the structures should be possible using fluorine based chemistry, but one needs to be aware that fluorine can also etch the substrate material.

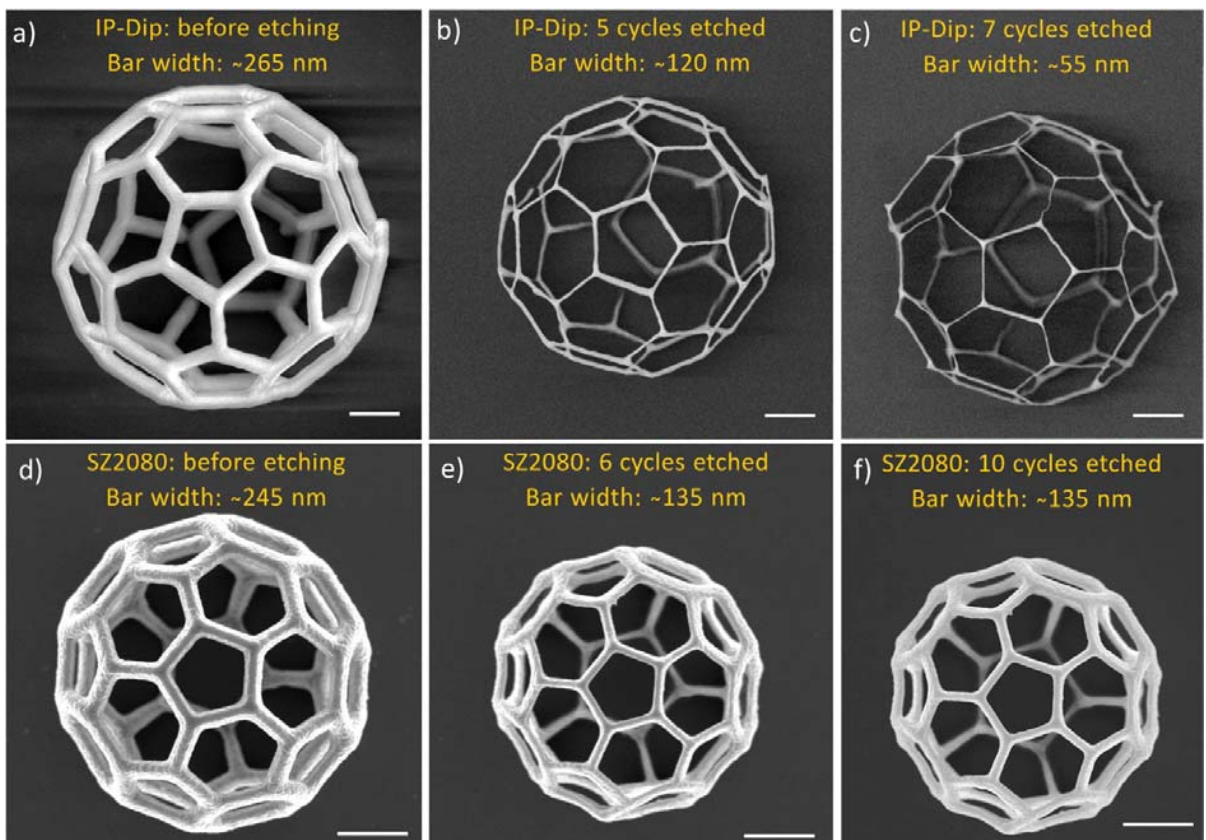

Figure 5: Isotropic oxygen plasma etching of polymerized 3D structures: a) initial buckyball made using IP-Dip resist, b) and c) – the same buckyball as in a) after 5 and 7 etching cycles, respectively. d) as polymerized SZ2080 structure, e) and f) – the same SZ 2080 structure after 6 and 10 processing cycles. Scale bars represent 1 μm in all images.

Table 1. Outcomes of the applied post processing steps

| Material \ Applied process | IP-Dip | SZ2080 |
|---|---|---|
| Pyrolysis at 690 °C | Shrinkage to 20-25% of initial size | Shrinkage to 60-70% of initial size |
| Isotropic oxygen plasma etching | Etching rate 15 ± 3 nm per min | Etching rate highly depends on the etching cycle number. The rate decreases from 12 ± 3 nm per min for the first cycle to almost no observable etching after 6th cycle |

The smallest lateral dimensions achievable using post-processing steps highly depend on the geometry of 3D architecture. Structures tend to collapse when a certain width/thickness limit of their building blocks is reached. In the case of 7.0 ± 0.1 µm diameter buckyballs made out of crosslinked IP-Dip resin (Figure 5a) the smallest obtained lateral bar width was 55 ± 8 nm (Figure 5c). At this width the composing bars already became distorted, but the entire structure was still intact. However, the structure completely collapsed during a further etching cycle. For 3.0 ± 0.1 µm diameter buckyball structures this limit was at 25 ± 6 nm as shown in Figure 6. The response of investigated resists to the applied post processing steps are summarized in Table 1.

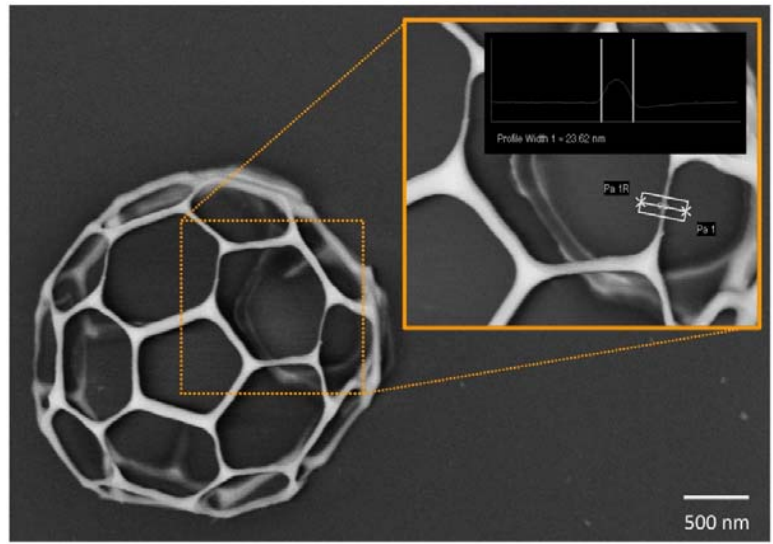

Figure 6: 3 µm diameter IP-Dip buckyball after 5 oxygen plasma etching cycles. Inset shows the obtained lateral width ~25 nm.

3.2. *Combination of Plasma Etching and Pyrolysis*

A post-processing combining both oxygen plasma etching and pyrolysis has also been investigated. Due to higher degree of shrinkage and better etching control IP-Dip resist was used to polymerize test structures. The composing bars of the structures were first thinned down in oxygen plasma from initial 340 ± 25 nm to ~220 ± 20 nm and then the entire buckyball structures were pyrolyzed in a tube furnace. By following this procedure we were able to precisely control the dimensions of the composing elements as well as the overall size of the structure. Figure 7 shows SEM images taken after each processing step and the final structure is presented in Figure 7c. The obtained structure is 1.60 ± 0.05 µm in diameter and the bars are 60 ± 10 nm wide. It is important to note that oxygen plasma etching does not inhibit further size reduction by pyrolysis as the thinned buckyball structures also shrunk to ~20% of their initial size.

The biggest advantage of the demonstrated post processing approach is that an aspect ratio of polymerized areas caused by the elongated laser spot can be avoided. Up to now, the highest resolution exposures were possible only as single pixel lines with typical thickness to width ratio of ~3. Otherwise, to produce 1:1 aspect ratio lines a few single pixel lines have to be written in an overlapped manner to increase the width, thus the resolution was reduced. In our case, it would be possible to write structures having 1:1 aspect ratio of their building blocks at the microscale and then use isotropic plasma etching and pyrolysis to thin and scale them down beyond 100 nm feature sizes. Pathways depicted in Figure 4 summarize available post processing routes to increase the resolution of laser polymerized structures.

Dry processing is another advantage of this method. Due to the wet development in 3D laser lithography, it is challenging to directly write nanoscale features as capillary forces tend to distort the structures even though critical point drying is used. This is not the case for plasma etching and pyrolysis that allow the limits of mechanical stability of the structures to be reached.

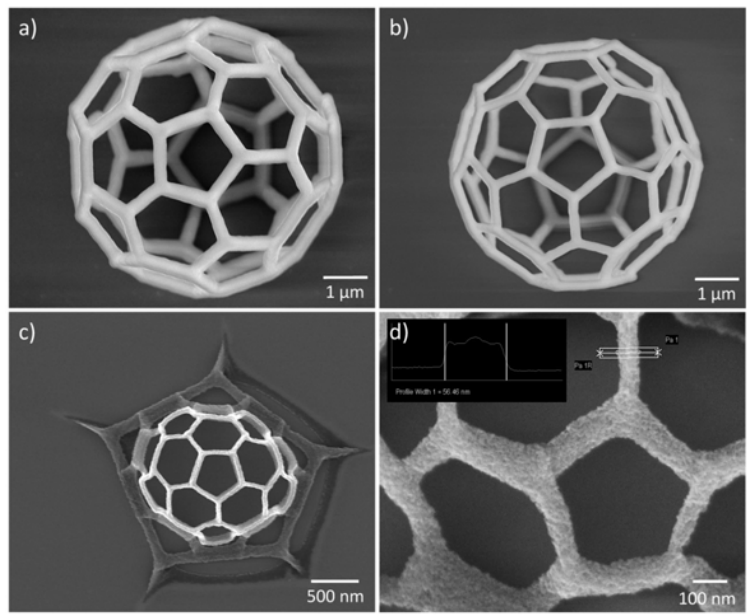

Figure 7: Combined post-processing of 7 µm diameter buckyball: a) initial structure produced from IP-Dip with lateral bar widths of ~340 nm; b) the same structure as in a) etched in oxygen plasma to reduce the bar widths to ~220 nm; c) structure obtained after the pyrolysis step, the diameter was reduced to ~1,6 µm; d) close up 45 degree tilt view demonstrating ~60 nm bar width.

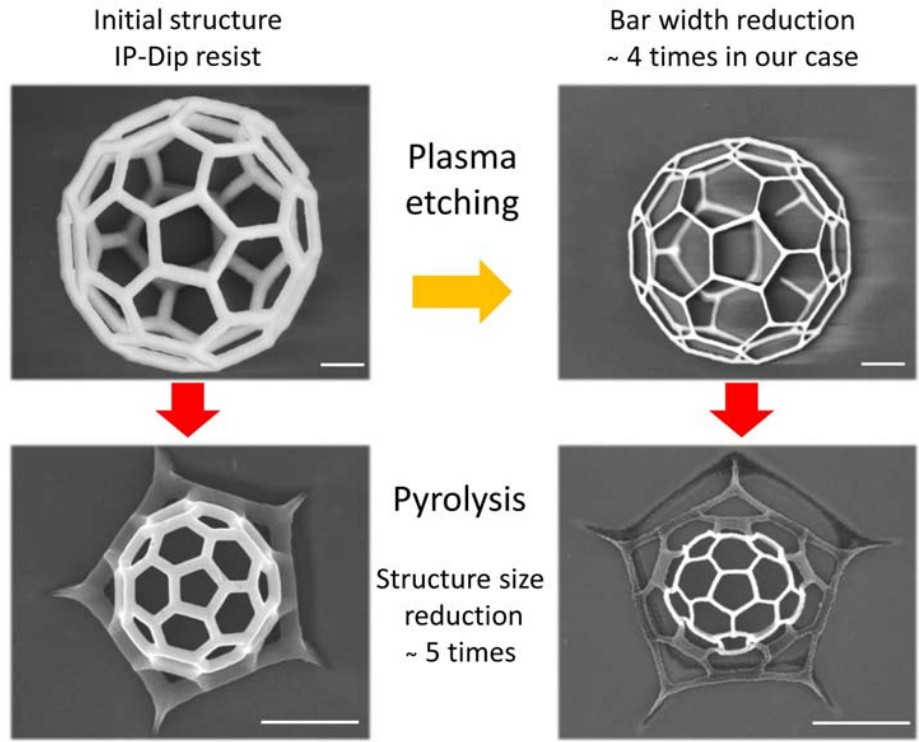

Figure 8: Possible post processing pathways to improve the resolution of 3D printed structures. Scale bars represent 1 μm in all images.

4. Conclusion

Dimensions of structural elements as well as size of the entire structure can be controlled at the nanoscale by combining isotropic plasma etching and pyrolysis as post processing steps. For 3D structures made from organic resists almost five fold scaling down was demonstrated. The smallest obtainable feature sizes were mostly limited by the mechanical stability of the polymerized resist and the geometry of 3D structure. Post processing not only leads to size reduction, but also material properties are changed due to thermal decomposition of the resist which typically becomes more robust, thus the mechanical stability of the structure is improved. Also, thermal down scaling gives an additional lever to engineer 3D nanostructures at the nanoscale by controlling shrinkage induced distortion pattern.

**Acknowledgments**

The research leading to these results has received funding from the EU-H2020 Research and Innovation program under Grant Agreement No. 654360 NFFA-Europe.

**Declaration of Interest**

All authors declare no conflict of interest.